# Terahertz spectroscopy of an electron-hole bilayer system in AlN/GaN/AlN quantum wells

H. Condori Quispe[1*], S.M. Islam[2*], S. Bader[2], A. Chanana[1], K. Lee[2], R. Chaudhuri[2], A. Nahata[1], H. G. Xing[2,3], D. Jena[2,3], and B. Sensale-Rodriguez[1]

[*]Equal Contribution

[1] Department of Electrical and Computer Engineering, University of Utah, Salt Lake City, UT 84112, USA

[2] School of Electrical and Computer Engineering, Cornell University, Ithaca, NY 14853, USA

[3] Department of Materials Science and Engineering, Cornell University, Ithaca, NY 14853, USA

**Abstract**

We describe studies on the nanoscale transport dynamics of carriers in strained AlN/GaN/AlN quantum wells: an electron-hole bilayer charge system with large difference in transport properties between the two charge layers. From electronic band diagram analysis, the presence of spatially separated two-dimensional electron and hole charge layers is predicted at opposite interfaces. Since these charge layers exhibit distinct spectral signatures at terahertz frequencies, a combination of terahertz and far-infrared spectroscopy enables us to extract (*a*) individual contributions to the total conductivity, as well as (*b*) effective scattering rates for charge-carriers in each layer. Furthermore, by comparing direct-current and terahertz extracted conductivity levels, we are able to determine the extent to which structural defects affect charge transport. Our results evidence that (*i*) a non-unity Hall-factor and (*ii*) the considerable contribution of holes to the overall conductivity, lead to a lower apparent mobility in Hall-effect measurements. Overall, our work demonstrates that terahertz spectroscopy is a suitable technique for the study of bilayer charge systems with large differences in transport properties between layers, such as quantum wells in III-Nitride semiconductors.

III-V compound semiconductors based on gallium nitride (GaN) have emerged as one of the most attractive materials for power electronics owing to its large bandgap (~3.4 eV), breakdown field (~3 MV/cm), and ability to operate at high frequencies, and, thus, remain the subject of intense research over the past several decades[1-10]. GaN showcases excellent properties such as: large peak electron velocity (~$2.5\times10^7$ cm/s), large saturation velocity (~$2.0\times10^7$ cm/s), and excellent carrier transport properties as evidenced by a high electron mobility ($\mu > 2,400$ $cm^2$/V.s) in GaN two-dimensional electron gases (2DEGs); these properties make GaN suitable for use as channel material in high frequency power devices[11-13]. In addition, aluminum nitride (AlN) withstands high temperature and high power density, owing to its very large band gap (~6.2 eV) and high thermal conductivity (~340 W/m.K), among other properties [14-17]. AlN/GaN/AlN quantum wells (QWs), i.e. thin GaN layers surrounded by AlN buffer and barrier layers, bring together all the above-mentioned features as an attractive platform for electronic applications. In this regard, QW devices can provide for tight electrostatic control and quantum confinement of charge carriers, thereby preventing degradation in performance by short-channel effects. Moreover, the superior thermal conductivity of AlN enables excellent heat dissipation.

A series of recent reports experimentally demonstrated the feasibility of ultra-thin GaN QWs surrounded by AlN buffer and barrier layers[18-24]. The discontinuity in spontaneous and piezoelectric polarization at the two heterojunctions of the QW gives rise to highly charged regions near the boundaries. Mobile carriers concentrate at these interfaces forming two-dimensional (2D) charge gases. Simulations predict that a population of distinct carrier species exists at each interface, i.e. in the upper interface a 2DEG is formed whereas at the bottom interface a 2D hole gas (2DHG) is formed[18-19]. Often, electrical characterization of these complex electron-hole bilayer systems is performed via Hall-effect measurements. While conventional Hall-effect measurements provide for a weighted average of the electron-hole bilayer system mobility and charge concentration, a more careful transport analysis is required to identify the individual effects of each type of carrier. In this work, we are able to decouple the electron and hole transport properties in this system by means of non-contact terahertz spectroscopy. A unique aspect of this approach is that the nanoscale transport properties, i.e. *conductivity and scattering rates*, for electron and hole layers can be independently extracted; this results from the distinct spectral features arising from each carrier species upon interaction with terahertz electromagnetic radiation. Our results evidence that the relatively modest electron mobility extracted from Hall-effect measurements ($\mu < 400$ cm$^2$/V.s) is not the result of structural defects such as dislocations, but likely the consequence of (*i*) a non-unity Hall-factor and (*ii*) the effect of holes, which due to their considerable contribution to the overall conductivity, substantially reduce the apparent Hall mobility. In effect, relaxation times of the 2DEGs associated with a drift mobility ~1,000 cm$^2$/V.s are consistently observed across all the analyzed samples. Overall, our results demonstrate an alternative, simple, yet insightful method to probe for the carrier transport properties in this bilayer system, which enables us to characterize (*a*) *the contributions of electron and hole layers to the overall conductivity*, (*b*) *the effective scattering rates for electrons and holes*, and (*c*) *the extent to which structural effects affect the overall conductivity*.

The schematic cross section of the analyzed hetero-structures is shown in **Fig. 1(a)**, which consists of a thick ~270 nm unintentionally doped AlN buffer layer, followed by a ~30 nm GaN layer to host the 2DEG channel, a 6 nm AlN barrier, and a 2 nm GaN passivation cap layer. Three different substrates were employed to support the QW structures and can be grouped into three sets: *Sample Set #1* utilized a 6H SiC substrate with a thickness $d_{S1} = 375$ μm; *Sample Set #2* utilized a AlN template with a thickness $d_{S2a} = 1$ μm on a sapphire substrate with a thickness $d_{S2b}$

= 430 μm; and *Sample Set #3* utilized a bulk AlN substrate of $d_{S3}$ = 550 μm. In all three sample sets, the QW heterostructures were grown using a Veeco Gen 930 plasma-assisted MBE system. Additional details about the growth conditions are discussed elsewhere[18, 21]. The schematic cross section of the QWs was examined by transmission electron microscopy (TEM). **Figure 1(b)** shows the cross-sectional TEM image of a representative GaN QW heterostructure. The carrier concentration and mobility were obtained from Hall-effect measurements across all samples; the extracted electrical transport properties are listed on the first three columns of **Table I**. All measurements reported in this work were performed at room temperature. The four samples in *Sample Set #1*, corresponding to different nucleation conditions of the buffer AlN layers such as Nitrogen rich (*S1a*), migration enhanced epitaxy (*S1b*), Gallium surfactant mediated (*S1c*), and short period super-lattice (*S1d*) growth conditions exhibited Hall-effect mobilities ($\mu_{Hall}$) of 366, 250, 359, and 290 cm$^2$/V.s, respectively, with charge densities ($n_s$) ranging between 2.81 to 4.78×10$^{13}$ cm$^{-2}$. The samples in *Sample Set #2* and *#3* showed charge densities of 3.20 and 4.31×10$^{13}$ cm$^{-2}$ but lower Hall-effect mobilities of 94 and 153 cm$^2$/V.s, respectively. We attribute the lower mobility observed in *Sample Sets #2* and *#3* to defects generated during un-optimized nucleation on the AlN surfaces[18].

In order to provide insight on the carrier distributions in the QWs, we numerically calculated the energy band diagrams for the active region, as shown in **Fig. 2(a)**. The electronic structure of the epitaxial layers was calculated using NEXTNANO$^3$ [25]. The program employs a self-consistent iterative procedure to solve the Schrodinger-Poisson equations for electron carriers and a self-consistent six-band *k.p* model for the valence band hole carriers[26-27]. The electron occupation is found to be limited to a single-subband despite the large charge density[28]; this is because of the large polarization fields. As shown in the inset of **Fig. 2(a)**, the electron inter-subband spacing (~0.4 eV) is larger than what is usually observed in typical AlGaN/GaN high-electron-mobility transistor (HEMT) structures, i.e. ~0.1 eV, where reports indicate that occupation of two subbands is often the case[29-30]. From this perspective we estimate that >95 % of the electron population arises from the first subband at room temperature, and nearly 100 % at low temperatures.

In the case of the valence band the situation requires a more careful transport analysis. **Fig. 2(b)** shows the hole gas in-plane energy dispersion formed by alternate heavy-hole (HH) and light-hole (LH) subbands. The first two subbands, which are depicted in the plot, are heavily occupied. For clarity, only one spin eigen-level is plotted for each subband. Spin-splitting due to structural inversion asymmetry is secondary to this analysis and the bulk-inversion asymmetry term [27] was neglected as is often done in inhomogeneous $k.p$ analysis[31]. The GaN QW is assumed to be strained pseudomorphically to the AlN substrate. While other reports have shown a relaxed GaN QW for thicknesses above 30 nm [32], the GaN QW is expected to be strained in the first several nm above the bottom interface, i.e. where the 2DHG is confined as depicted in **Fig 2(a)**. Compressive strain in the GaN QW shifts the split-off (SO) band hundreds of meV below the HH and LH bands, as shown in the inset of **Fig. 2(b)**. As in the bulk case, the first two sub-bands (one HH and one LH) are nearly degenerate at $k_t = 0$; the populations of carriers are proportional to the density of states, which in turn is proportional to the transverse effective mass ($m_\perp$) for a two-dimensional hole gas. Meanwhile, the mobility, under a low field approximation, is inversely proportional to the longitudinal effective mass ($m_\parallel$). Since the conductivity contribution for each population is the product of carrier concentration ($\propto m_\perp$) and carrier mobility ($\propto 1/m_\parallel$) and in GaN considered in this work the effective mass is isotropic ($m_\parallel \sim m_\perp$), their dependencies on effective mass cancel. As a result, we will later consider that each band population (HH and LH) contributes nearly equally to the measured hole conductivity when experimentally extracting hole densities.

To model the effect of free carrier dynamics on the terahertz transmission through the samples, the 2DEG/2DHG conductivities were modeled using a Drude frequency dispersion [33-35]. This assumption is further validated via measurements through GaN control samples containing just one charge carrier species, i.e. only electrons or holes. Our results evidence that both electrons and holes are indeed characterized by Drude responses with single scattering times; see *Supplementary Material*. Since the thickness of the QW is negligible compared to the wavelength of the terahertz radiation, the system can be modeled via an effective (bilayer) conductivity given by the sum of the electron and hole layer conductivities [36]:

$$\sigma_b(\omega) = \sigma_e^0/(1 + j\omega\tau_e) + \sigma_h^0/(1 + j\omega\tau_h), \qquad (1)$$

where, $\sigma_b(\omega)$ is the bilayer dynamic conductivity, $\sigma_e^0$ is the zero-frequency dynamic conductivity of the 2DEG, $\tau_e$ is the momentum relaxation time for electrons in the 2DEG, $\sigma_h^0$ is the zero-frequency dynamic conductivity of the 2DEG, $\tau_h$ is the 2DHG momentum relaxation time, and $\omega$ is angular frequency. **Figure 2(c)** shows the calculated bilayer conductivity following Eq. (1) and assuming 2DEG and 2DHG direct-current (DC) conductivities of 1 and 0.1 mS and momentum relaxation times of 160 and 10 fs, respectively, which correspond to typical values reported in the literature[18, 37]. These relaxation times are in the range of those observed in the control samples (see *Supplementary Material*, **Fig. S1**). As depicted in **Fig. 2(c)** two distinct frequency windows can be identified: (*i*) below ~3 THz, where both electrons and holes contribute to the overall conductivity, and (*ii*) above ~3 THz, which is beyond the 2DEG Drude roll-off and thus the overall conductivity is set mainly by the 2DHG.

We used two different terahertz systems to characterize the samples: a continuous wave (CW) terahertz spectrometer and a terahertz time-domain spectrometer (THz-TDS). The CW spectrometer (TOPTICA Photonics) used InGaAs photo-mixers at 1550 nm for both generation and detection. In the THz-TDS setup, a broadband terahertz pulse was generated via optical rectification using a <110> ZnTe crystal. The sample was placed at the focal plane of the terahertz beam, and its response was modulated using electro-optic sampling in a separate <110> ZnTe detection crystal[38]. The transmitted signal was Fourier transformed to obtain its frequency spectra, which was then normalized by the response of a reference substrate. Data in the 0.4 to 1.6 THz frequency range was obtained from THz-TDS measurements; CW transmission measurements (0.3 to 0.6 THz) were performed so to confirm the low frequency end of the THz-TDS data. The effective bilayer conductivity of the electron/hole system in the GaN QW was experimentally determined by fitting the normalized terahertz transmission ($T/T_0$) using the following expression[39]:

$$T/T_0 = \left|1 + Z_0 \sigma_b(\omega)/(1 + \sqrt{\varepsilon_{sub}})\right|^{-2}, \quad (2)$$

where the transmission ($T$) through each sample is normalized by that through an appropriate bare reference substrate ($T_0$); $Z_0$ is the vacuum impedance ($Z_0=377\Omega$), and $\varepsilon_{sub}$ is the relative permittivity of the substrate, which corresponds to 9.7, 9.3, and 9.1 for samples in *Set* #1 (SiC), #2 (Sapphire), and #3 (AlN), respectively. From Eqns. (1) and (2) the transmission spectrum is expected to exhibit unique spectral signatures arising from the interactions of terahertz waves with different charge carrier species. As shown in **Fig. 2(c)** electrons and holes depict very distinct spectral behaviors. As a result, two key spectral features are anticipated: (*i*) a rapid extinction drop owing to the long relaxation time characteristic of electrons, and (*ii*) a gradual decay as a result of the very short relaxation time of holes.

We can extract several material parameters from the transmission data in the THz-TDS frequency window. By assuming that $\omega\tau_h \ll 1$, which is the case in the 0.4 to 1.6 THz TDS range of operation, Eq. (1) can be reduced to:

$$\sigma_b(\omega) = (\sigma_b^0 - \sigma_h^0)/(1 + j\omega\tau_e) + \sigma_h^0, \tag{3}$$

Therefore: $\sigma_b^0$, $\sigma_h^0$, and $\tau_e$ can be extracted by fitting to the model in Eqns. (2) and (3). In **Fig. 3(a),** we show the transmission spectra through samples *S1d*, *S2*, and *S3*, as well as its fitting to the model. The extracted bilayer conductivity ($\sigma_b^0 = \sigma_h^0 + \sigma_e^0$) is 1.51 ± 0.07, 0.90 ± 0.15, and 1.88 ± 0.16 mS for samples *S1d*, *S2*, and *S3* respectively. Moreover, the extracted electron momentum relaxation times are 122.0 ± 10.66, 100.85 ± 4.5, and 94.28 ± 27.37 fs, respectively. Listed in **Table I** are the extracted values of $\sigma_b^0$, $\sigma_h^0$, and $\tau_e$ for all the analyzed samples. *Because of the limited spectral range of these measurements, a large uncertainty is observed in the extracted hole zero-frequency dynamic conductivity.* Although terahertz spectroscopy enables us to directly extract the conductivity and relaxation times associated with charge carriers, it is also possible to indirectly extract other parameters, such as charge density and mobility by assuming an appropriate effective mass. By assuming an electron effective mass of $m_e = 0.2m_0$, electron drift mobilities ranging from 828 ± 240 (*Sample S3*) to 1201 ± 31 cm$^2$/V.s (*Sample S1a*) are extracted across all the analyzed samples. Comparison of these mobility levels with those extracted from Hall measurements requires careful analysis and will be discussed later in the manuscript. Moreover, the corresponding electron densities range from 0.6 ± 0.1 (*Sample S2*) to 1.3 ± 0.1 ×10$^{13}$ cm$^{-2}$ (*Sample S1b*).

To further validate the measured data, we performed CW terahertz measurements. A representative transmission spectra, i.e. for sample *S1d*, is depicted in **Fig. 3(b)**. Owing to the presence of Fabry-Perot resonances, the experimental transmission data was fitted to an analytical model based on the transfer-matrix formalism following the methods described in Ref. [40]. In **Fig. 3(b)**, we show the calculated transmission spectra that best fits the experimental data, the extracted zero-frequency bilayer conductivity is 1.48 ± 0.12 mS. This value agrees well with the value obtained from THz-TDS. Overall an excellent agreement was observed between THz-TDS and CW measurements across all samples.

To qualitatively compare the terahertz extracted zero-frequency bilayer conductivity ($\sigma_b^0$) with the bilayer DC conductivity obtained from Hall-effect measurements ($\sigma_b^{Hall}$), we computed their ratio ($R=\sigma_b^0/\sigma_b^{Hall}$). **Figure 3(c)** depicts the computed $R$ for all the analyzed samples. In general we observe that for samples in *Sample Set #1, R* is close to unity. However, for samples corresponding to *Sample Sets #2 and #3*, the terahertz-extracted conductivities are up to two-times larger than those extracted from Hall-effect measurements. In this regard, it is worth mentioning that terahertz spectroscopy at a frequency ω probes charge transport in a characteristic length given by $L(\omega) = \sqrt{D/\omega}$, which corresponds to the length-scale that a gas of carriers diffuses before the terahertz electromagnetic fields reverse direction[41]. As a result $L(\omega)$ depends on the probe frequency as well as on the diffusion coefficient (*D*) for charge carriers in the material under test, which ranges in the order of 5-30 cm$^2$/s for charge carriers in GaN[42]. From this perspective, we find that at terahertz frequencies, i.e. *f* > 300 GHz, the characteristic length is <40 nm. The scale of this characteristic length indicates that the zero-frequency bilayer conductivity extracted from terahertz measurements is a spatially-averaged nanoscale conductivity and is thus less affected by microscopic scale effects with a characteristic scattering length >40 nm than the conductivity extracted from DC Hall-effect measurements. In this regard, the fact that the largest *R* occurs in samples exhibiting the lowest Hall-effect mobility (i.e. *Sample Sets #2 and #3*) suggests that defects and dislocations present in these sample sets introduce additional scattering with a characteristics length >40 nm, thus heavily affecting the DC charge transport in these particular samples. This is consistent with previous observations in other materials systems[43-44]. On the other hand, in samples exhibiting the highest Hall-effect mobilities (i.e. *Sample Set #1*) and a unity *R*, transport is likely limited by other factors with a characteristic length <40 nm such as interface roughness scattering, phonon scattering, Stark-effect scattering, interlayer Coulomb drag effects, etc., as discussed in[18].

Extraction of the relaxation time for the 2DHG requires extending our measurements beyond the THz-TDS frequency window. For this purpose we performed transmission measurements in the 3 to 14 THz frequency range employing an FTIR system (Bruker IFS-88). Owing to their large size as well as low substrate surface roughness, two samples from *Sample Set #1* (*S1c* and *S1d*) were analyzed. A representative extinction spectrum (sample *S1d*) is depicted in **Fig. 4**, which consists of THz-TDS data (0.4 to 1.6 THz) and FTIR data (3 to 14 THz). The upper frequency limit for the FTIR measurement was set to 14 THz in order to remain below the Restrahlen band corresponding to TO phonons in SiC[45]. The extinction spectra were fitted to the model in Eqns. (1) and (2), from which $\sigma_e^0$, $\tau_e$, $\sigma_h^0$, and $\tau_h$ were extracted. Overall the measured extinction spectra closely agree with this model as observed in **Fig. 4**. Furthermore, two different frequency regimes are distinguished, which are correspondingly associated with a short and a long scattering time. The fact that the extinction does not reach zero even at the upper frequency of 14 THz is a signature of free carrier absorption by holes in the 2DHG located at the bottom interface of the QW. The extracted zero-frequency electron conductivity is: 1.72 ± 0.04 and 1.32 ± 0.08 mS for samples *S1c* and *S1d*, respectively. The extracted zero-frequency hole conductivities are 0.11 ± 0.05 and 0.19 ± 0.04 mS. Moreover, the extracted electron momentum relaxation times are

155.16 ± 7.44 and 124.72 ± 9.18 fs, and the extracted hole relaxation times are 7.40 ± 4.23 and 19.20 ± 2.81 fs, respectively. *A close agreement is observed between these values, which are obtained from a fitting over an extended frequency spectra, and those extracted only from THz-TDS data.* It is also worth mentioning that the extracted hole momentum relaxation times are in good agreement with those obtained from FTIR spectroscopy of *p*-GaN samples on semi-insulating GaN/SiC (11.8 ± 4.2 fs); see *Supplementary Material*, **Fig. S1**.

From the energy band diagram simulations shown in **Fig. 2(b)**, the estimated hole populations in the hole subbands are $4.0 \times 10^{13}$ cm$^{-2}$ and $0.7 \times 10^{13}$ cm$^{-2}$ for HH and LH, respectively; the effective masses for HH and LH correspond to $m_{HH} = 2.11 m_0$ and $m_{LH} = 0.41 m_0$, respectively. By assuming that carriers in both subbands are subjected to similar scattering mechanisms and thus considering similar scattering times for both populations, it is expected that each population will contribute equally to the measured terahertz extinction. Apportioning half of the conductivity to each population, and employing the calculated effective mass for each subband, the following hole densities are extracted from the experimental data: $5.51 \pm 2.41 \times 10^{13}$ cm$^{-2}$ (HH) & $1.07 \pm 0.38 \times 10^{13}$ cm$^{-2}$ (LH), and $3.67 \pm 0.79 \times 10^{13}$ cm$^{-2}$ (HH) & $0.71 \pm 0.15 \times 10^{13}$ cm$^{-2}$ (LH), for samples *S1c* and *S1d*, respectively, which are in good agreement with the values predicted from band diagram simulations. Furthermore, from the extracted relaxation times, the following hole mobilities are calculated: $6.17 \pm 3.52$ cm$^2$/V.s (HH) & $31.70 \pm 18.12$ cm$^2$/V.s (LH), and $15.98 \pm 2.34$ cm$^2$/V.s (HH) & $82.25 \pm 12$ cm$^2$/V.s (LH), for samples *S1c* and *S1d*, respectively.

At this end, since the full-transport properties for the 2DEG and the 2DHG have been extracted, it is possible for us to explain physical reasons behind the modest observed Hall mobility values. This will be attributed to two effects: (*i*) a non-unity Hall-factor and (*ii*) the effect of holes, which due to their considerable contribution to the overall conductivity, substantially reduce the apparent Hall mobility. In general, for a multilayer system containing multiple electron and hole layers, the resultant Hall mobility is a weighted average given by:

$$\mu_{Hall} = \sum_{i\ (electrons)} \mu_i \frac{\sigma_i}{\sigma_{total}} - \sum_{j\ (holes)} \mu_j \frac{\sigma_j}{\sigma_{total}} \quad (4)$$

Across all the analyzed samples we observe an average electron density of ~$1.0\times10^{13}$ cm$^{-2}$, an average electron drift mobility of ~1,000 cm$^2$/V.s, and an overall contribution of holes to the total conductivity of up to 20%. Using these numbers, and assuming LH and HH mobilities of ~60 cm$^2$/V.s and ~10 cm$^2$/V.s, respectively, we estimate a weighted mobility of ~790 cm$^2$/V.s, as per Eq. (4). From this perspective, *the smaller Hall-effect mobility observed in experiments is in part a result of holes contributing to the overall conductivity and reducing the apparent Hall mobility.* However, by only accounting for this effect, there is still a significant difference between THz-extracted and Hall-effect extracted mobility levels. For many semiconductors, including GaN, a conversion factor (Hall-factor) is defined to obtain the true carrier concentration from the measured Hall coefficient[46-47]. This Hall-factor also represents the ratio between Hall mobility and drift mobility ($r_{Hall} = \mu_{drift} / \mu_{Hall}$)[48]. An issue when determining the Hall factor is that it is typically not possible to independently extract drift and Hall mobilities under same experimental conditions. From this point of view extraction of $r_{Hall}$ often relies on fitting of the Hall mobility versus temperature to theoretical models. However, these methods are adequate at low temperatures and low charge concentrations, where elastic and isotropic scattering processes are dominant[48]. In spite of these limitations, non-unity Hall-factors have been widely reported in the literature, with both $r_{Hall} > 1$ and $r_{Hall} < 1$ depending on dominating scattering mechanisms in the samples[48-51]. In this work our methodology enabled us to independently extract both $\mu_{Hall}$ and $\mu_{drift}$ under the same experimental conditions. Based on the estimated weighted mobility, we calculate $r_{Hall}$ ~ 0.5 in *Sample Set #1*, which has higher mobilities. This value is consistent with our observations in AlGaN/GaN 2DEGs (see *Supplementary Material*) where a Hall mobility of ~1,400 cm$^2$/V.s was observed in spite of a much larger THz-extracted drift mobility of 2,000 cm$^2$/V.s (i.e. $r_{Hall}$ ~ 0.7). Investigations on the physical reasons behind these particular extracted values fall out of the scope of this manuscript and will be the subject of future investigations.

In conclusion, we have reported on terahertz properties of strained AlN/GaN/AlN QWs. Simulations predict that carriers concentrate near the interfaces forming two-dimensional charge gases of distinct carrier species. From THz-TDS measurements, we were able to extract an effective zero-frequency bilayer conductivity in a non-contact manner. When compared to the measured DC electrical conductivity, the THz approach provides valuable information on the extent to which charge transport is affected by scattering of varied characteristic lengths. Using a combination of THz-TDS and FTIR measurements and assuming Drude models for both charge species, we were able to individually extract their nanoscale transport properties, i.e. conductivity as well as scattering rates. Moreover, our results evidence that a non-unity Hall-factor and the considerable contribution of holes to the overall conductivity, lead to a reduced apparent electron mobility obtained from Hall-effect measurements.

**Supplementary Material**

See *Supplementary Material* for terahertz spectroscopy data of control samples consisting of (*i*) an AlGaN/GaN HEMT epitaxial structure, and (*ii*) a *p*-GaN film.

**Acknowledgement**

This work was in part supported by the Office of Naval Research MURI, N00014-11-1-0721, monitored by Paul Maki and by the Air Force Office of Scientific Research (AFOSR), FA9550-17-1-0048, monitored by Ken Goretta. This work was also supported by the NSF MRSEC program at the University of Utah under Grant No. #DMR 1121252 and by NSF ECCS #1407959.

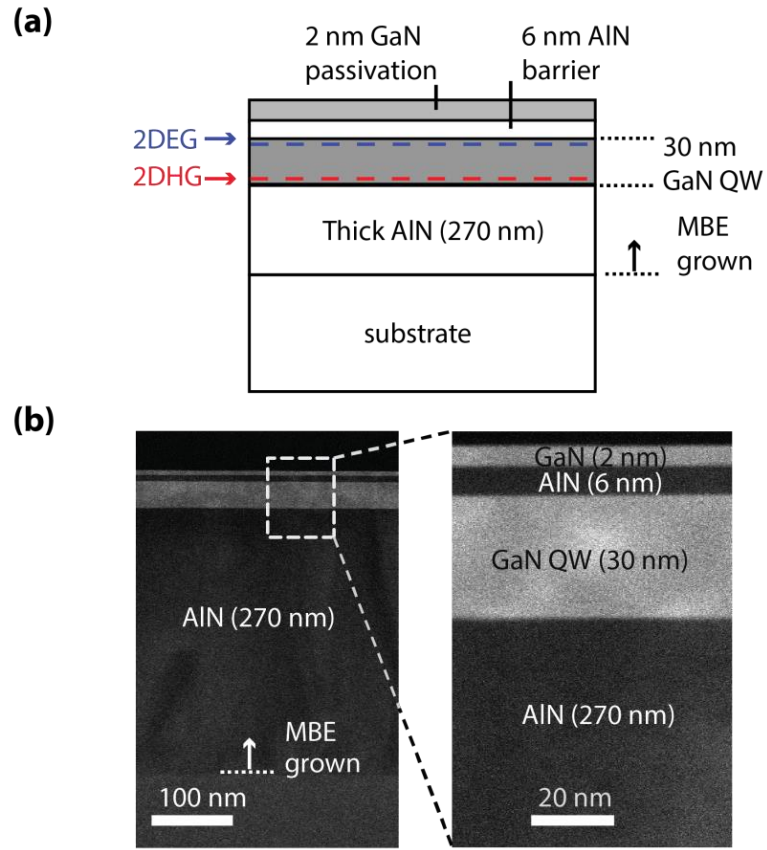

**Figure 1. (a)** Schematic cross section of the analyzed AlN/GaN/AlN quantum well structures. **(b)** Transmission electron microscopy (TEM) of a representative GaN QW under study.

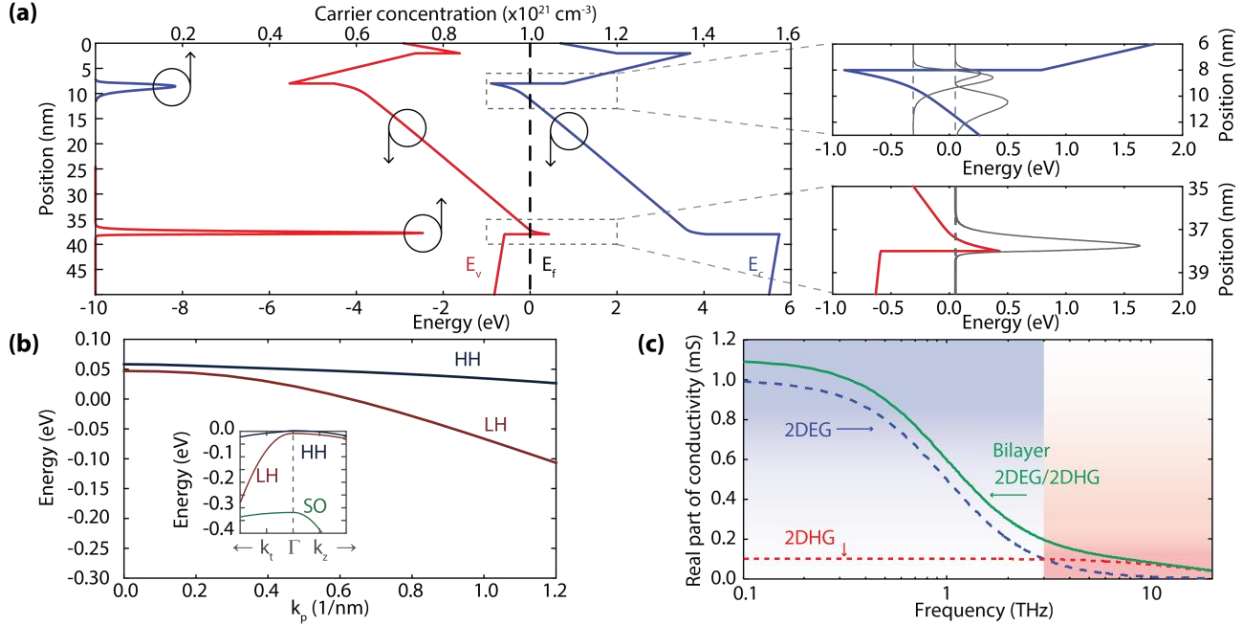

**Figure 2. (a)** Self-consistent Schrodinger-Poisson 6x6 multiband-$k.p$ simulation of the electronic structure of the AlN/GaN/AlN QW and calculated charge concentrations at the interfaces. Strong internal polarization fields induce 2DEG and 2DHG layers at the upper and lower interfaces of the well, respectively. Electron and hole wave-functions and energy levels are depicted in the insets. **(b)** The hole-gas first two subbands corresponding to heavy-holes (HH) and light-holes (LH); the inset depicts the bulk strained valence bands along with the split-off (SO) band. The Fermi level is set at $E_f = 0$ eV. Note: for clarity, only one spin eigenlevel is plotted for each subband. **(c)** Modeled effective bilayer dynamic conductivity for the 2DEG/2DHG system as per Eq. (1). The 2DEG and 2DHG conductivities were modeled using a Drude model with zero-frequency conductivities of 1 & 0.1 mS and momentum relaxation times of 160 & 10 fs, respectively.

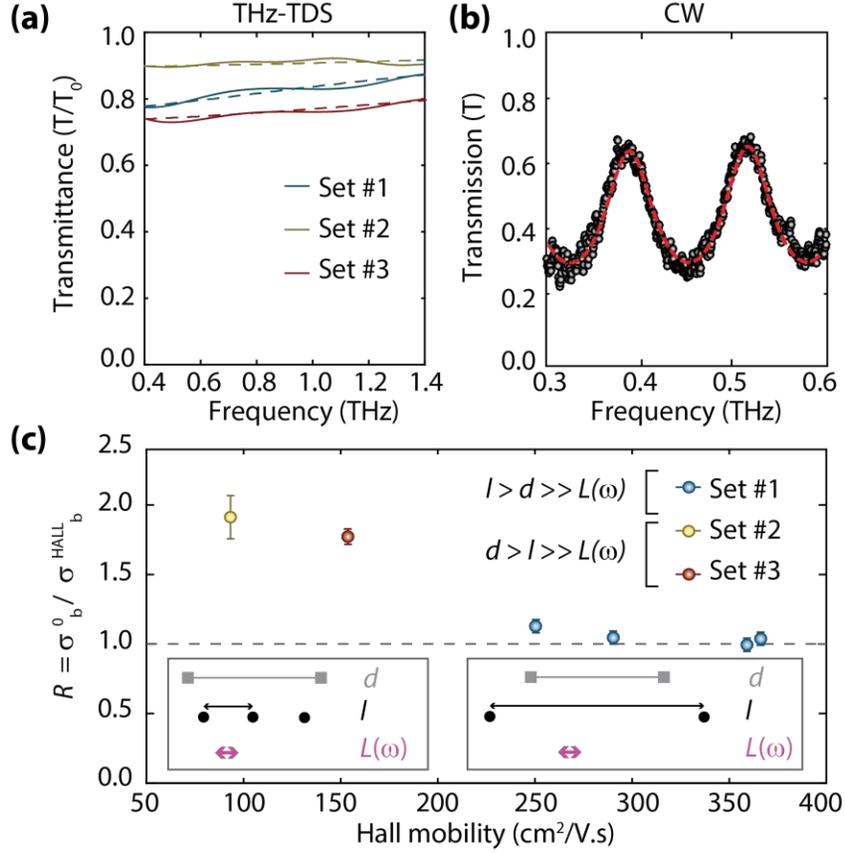

**Figure 3. (a)** TDS measurement (continuous lines) and its fit to the model given by Eqns. (2) and (3) (dashed lines) for three representative samples corresponding to *Sample Sets #1*, *#2*, and *#3*. **(b)** CW terahertz measurement and its fit for a representative sample corresponding to *Sample Set #1*. **(c)** Ratio ($R$) between the terahertz-extracted zero-frequency bilayer conductivity ($\sigma_b^0$) and the bilayer DC conductivity obtained from Hall-effect measurements ($\sigma_b^{Hall}$) versus Hall-effect mobility for all the analyzed samples. Depicted in the insets are illustrative sketches of the relations between $L(\omega)$, $l$, and $d$ for *Sample Set #1* (right inset) and *Sample Sets #2* and *#3* (left inset), where $L(\omega)$ is the characteristic length at which transport is probed by terahertz spectroscopy, $d$ is the characteristic length at which transport is probed in DC measurements, and $l$ is the mean free path between scattering events due to structural effects. The fact that $R > 1$ in *Sample Sets #2* and *#3* is attributed to $d$, $l$, and $L(\omega)$ satisfying $d > l > L(\omega)$.

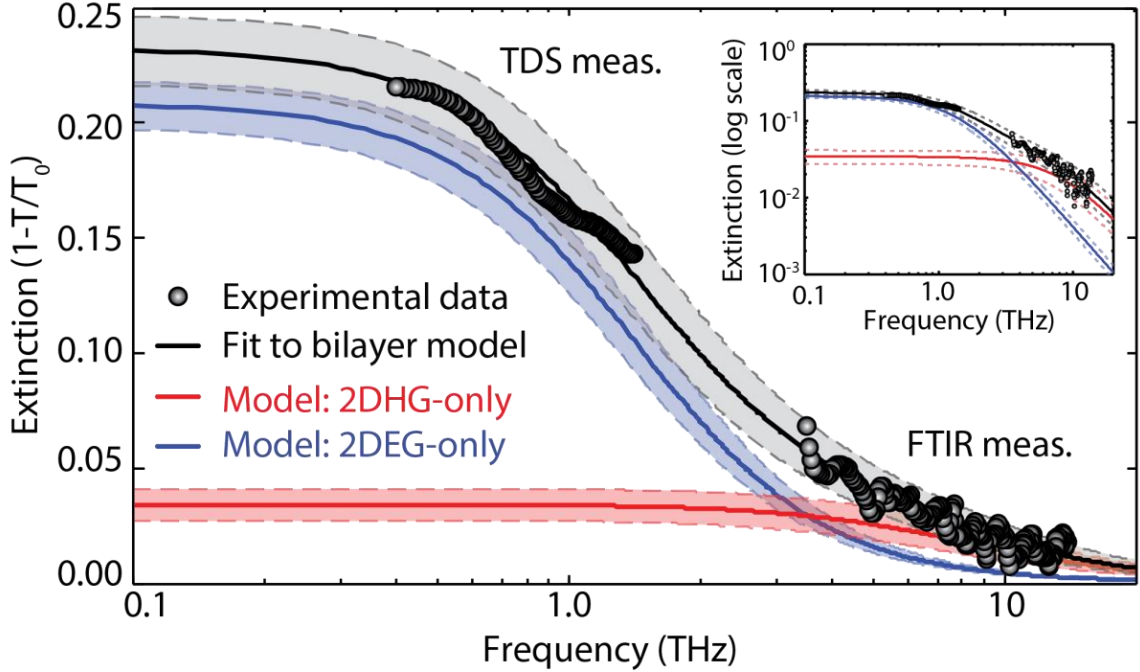

**Figure 4.** Representative extinction spectra over an extended frequency range (*Sample #1d*). The data from 0.4 to 1.4 THz was obtained via THz-TDS, while the data from 3 to 14 THz was obtained via FTIR spectroscopy. The continuous black line corresponds to the best fit of the measured data to the analytical model given by Eqns. (1) and (2). For reference: modeled spectral signatures corresponding to a 2DEG-only (blue) and a 2DHG-only (red) exhibiting zero-frequency conductivities and relaxation times consistent with those experimentally extracted for this sample are also shown in the plot. The colored shaded regions located between dashed traces represent uncertainty margins at the 95 % confidence level. The inset depicts the same plot in log-log scale, showing in more detail the fit of the high frequency data to the model.

**Table I.** Extracted transport properties from Hall-effect measurements and THz-TDS.

| Sample | Hall-effect measurements | | | THz-TDS | | | |
|---|---|---|---|---|---|---|---|
| | $\mu_{Hall}$ (cm$^2$/V.s) | $n_s$ (10$^{13}$ cm$^{-2}$) | $\sigma_b^{Hall}$ (mS) | $\sigma_b^0$ (mS) | $\sigma_h^0$ (mS) | $\tau_e$ (fs) | $\mu_{e,drift}$ (cm$^2$/V.s) |
| S1a | 366 | 2.81 | 1.65 | 1.78 ±0.09 | 0.13 ±0.08 | 136.80 ±3.54 | 1201 ±31 |
| S1b | 250 | 4.78 | 1.91 | 2.15 ±0.04 | 0.21 ±0.12 | 105.70 ±6.41 | 928 ±56 |
| S1c | 359 | 3.24 | 1.86 | 1.85 ±0.08 | 0.18 ±0.10 | 134.44 ±4.78 | 1180 ±42 |
| S1d | 290 | 3.1 | 1.44 | 1.51 ±0.07 | 0.24 ±0.13 | 122.00 ±10.66 | 1071 ±94 |
| S2 | 93.7 | 3.2 | 0.47 | 0.90 ±0.15 | 0.08 ±0.03 | 100.85 ±4.50 | 886 ±40 |
| S3 | 153 | 4.31 | 1.05 | 1.88 ±0.16 | 0.17 ±0.09 | 94.28 ±27.37 | 828 ±240 |